\documentclass[prl,nofootinbib,preprint,groupedaddress,longbibliography]{revtex4-2}
\usepackage{natbib}
\usepackage{amsmath} 
\usepackage{amssymb}
\usepackage{graphicx}
\usepackage{color} 
\usepackage{url}

\usepackage{hyperref}

\begin{document}
\title{Nontrivial zeros of the Riemann zeta function on the celestial circle}
\author{Wei~Fan}
\email{fanwei@just.edu.cn}
\affiliation{Department of Physics, School of Science, Jiangsu University of Science and
Technology, Zhenjiang 212114, China}
\date{\today}
\begin{abstract}
In this short letter, we reformulate the Riemann zeta function using the holographic framework of the celestial conformal field theory. For spacetime dimension larger than our Minkowski spacetime $M^4$,  the Riemann zeta function is connected with the sum of the conformal primary wavefunctions evaluated over a chain of points on the holographic boundary. Using analytic continuation, it follows that the  nontrivial zeros of the Riemann zeta function is  connected with the scaling dimension of conformal operators on the celestial circle. We discuss possible considerations with the spectrum of the celestial conformal field theory, number theory and topology. 
\end{abstract}

\maketitle

The Riemann zeta function $\zeta(s)$ is a meromorphic function on the complex plane $\mathbb{C}$, defined by analytic continuation from the absolutely convergent series
\begin{equation}
    \zeta(s):=\sum_{n=1}^{\infty} \frac{1}{n^s}, \, \Re(s)>1.
\end{equation}
It has trivial zeros at negative even integers and is conjectured to have all its nontrivial zeros lying in the line $\Re(s)=1/2$, which is the famous Riemann hypothesis~\cite{Bombieri2000ProblemsOT}. 
Here we will \emph{reformulate the zeros of $\zeta(s)$ on the celestial circle~\eqref{eq:zeroS}}, a special case of the celestial conformal field theory (CCFT). 

The CCFT is a recently proposed holography framework that connects many physics fields, like asymmptotic symmetries, scattering amplitudes and AdS/CFT. See~\cite{Strominger:2017zoo} for its origins and~\cite{Pasterski:2021raf} for its recent developments, and references therein. The bulk QFT in Minkowski spacetime $\mathbb{R}^{1,1+d}$ can be recast as a holographic CFT in the boundary $\mathbb{R}^{d}$. Due to causality, only physics within the light-cone is considered and the boundary is a compact subset of $\mathbb{R}^{d}$. In the case $d=2$ it is the famous celestial sphere  and in $d=1$, the boundary is called  the celestial circle~\cite{Lam:2017ofc}. For massless theory, the key ingredient of this holography is the conformal primary wavefunction~\cite{Pasterski:2017kqt} 
\begin{align}
 &\varphi_{\Delta}^{ \pm}\left(X^\mu ; \vec{w}\right) \equiv \int_0^{\infty} d \omega \omega^{\Delta-1} e^{ \pm i \omega q(\vec{w}) \cdot X-\epsilon \omega}\nonumber\\
 &=\frac{(\mp i)^{\Delta} \Gamma(\Delta)}{(-q(\vec{w}) \cdot X \mp i \epsilon)^{\Delta}},\quad  X\in \mathbb{R}^{1,1+d}, q(\vec{w})\in \mathbb{R}^{d},
\end{align}
where $q(\vec{w})$ is the asymptotic direction of the propagating massless particle and $\omega$ is its energy. The $\pm$ sign indicate the incoming/outgoing states and $\epsilon$ is a regulator for the integral that should be taken as zero $\epsilon\to 0^{+}$ in the end. This is a Mellin transform of the plane wave and the dual variable of energy $\omega$ is the scaling dimension $\Delta$ of a conformal operator that must be in the principal continuous series~\cite{Mack:1974jjo} defined as
\begin{equation}
    \Delta = \frac{d}{2} + i\lambda, \quad \lambda\in \mathbb{R}.
\end{equation}
So each massless particle has an associated conformal operator in CCFT. This conformal primary wavefunction is the bulk-to-boundary propagator~\cite{Cheung:2016iub,Witten:1998qj}, which are external lines in Witten diagrams that connect the bulk with the boundary within a hyperbolic slicing of the Minkowski spacetime.

Now for a given bulk point $X$, suppose that we can choose a chain of points on the boundary satisfying 
\begin{equation}
\label{eq:chain}
 \{q_n\in \mathbb{R}^{d}\mid  -q_n(\vec{w}) \cdot X = n,\quad  n\in \mathbb{Z}_{+} \}.
\end{equation}
Over this chain, let's define the  sum of  values of the conformal primary wavefunction or the bulk-to-boundary propagator as 
\begin{equation}
\label{eq:defSum}
S_{\varphi}(\Delta)\equiv \sum_{n=1}^{\infty} \varphi_{\Delta}^{ \pm}\left( -q_n(\vec{w}) \cdot X = n \right),     
\end{equation}
which is a function of $\Delta$ for a given $\varphi$. 
Then  for $d>2$, it evaluates to the Riemann zeta function $\zeta(\Delta=d/2 + i \lambda)$
\begin{align}
\label{eq:connectiond}
S_{\varphi}(\Delta)= \sum_{n=1}^{\infty} \frac{(\mp i)^{\Delta} \Gamma(\Delta)}{(n \mp i \epsilon)^{\Delta}}
\overset{\epsilon\to 0^+}{=} (\mp i)^{\Delta} \Gamma(\Delta) \zeta(\Delta), 
\end{align}
where in the last step the regulator is removed.

Like in   dimensional regularization~\cite{tHooft:1972tcz}, if we can do the  analytic continuation on the spacetime dimension from $d>2$ to $d=1$, the sum $S_{\varphi}(\Delta)$ will be connected with the line  $\Re(\Delta)=1/2$ of nontrivial zeros of $\zeta(\Delta)$
\begin{equation}
\label{eq:connectionLine}
  S_{\varphi}(\frac{1}{2}+i\lambda) = (\mp i)^{\frac{1}{2}+i\lambda} \Gamma(\frac{1}{2}+i\lambda) \zeta(\frac{1}{2}+i\lambda). 
\end{equation}
Because the gamma function is nonzero, we have 
\begin{equation}
\label{eq:zeroS}
 \zeta(\frac{1}{2}+i\lambda^*) =0 \cong  S_{\varphi}(\frac{1}{2}+i\lambda^*)=0,
\end{equation}
where the nontrivial zeros $1/2+i\lambda^*$ of $\zeta(\Delta)$ are equivalent to zeros of  $S_{\varphi}(\Delta)$. So on the chain~\eqref{eq:chain} of the celestial circle, the zeros of the sum of evaluated conformal primary wavefunctions are equivalent to the nontrivial zeros of the Riemann zeta function of the scaling dimension. 

Note that this is a strong constraint on both the scaling dimension $\Delta$ and the functional form $\varphi$ of the conformal primary wavefunction. So  the nontrivial zeros of   $\zeta(1/2+i\lambda)$ select a special class of conformal primary wavefunctions $\varphi_{\Delta}$ from the principal continuous series. This special class of selected $\varphi_{\Delta}$ determines a corresponding set of conformal operators on the celestial circle. \emph{What's the meaning of this set of conformal operators?} We know that in minimal models~\cite{Belavin:1984vu} of standard CFT, the spectrum is determined by coprime integers. Here the special set of conformal operators is selected by the nontrivial zeros of the Riemann zeta function,  which is related with the distribution of prime numbers. The similarity between these two situations suggests that the selected set of conformal operators might be the spectrum of the CCFT on the celestial circle. There might be a more fundamental connection with the number theory of pure mathematics, among the minimal models and the CCFT on the celestial circle. 

Our result~\eqref{eq:connectionLine}  depends on the validity of the analytic continuation from $d>2$ to $d=1$. It is a valid procedure if we only treat it as a mathematical technique applied on the function~\eqref{eq:connectiond}. But we are still faced with the question of \emph{what is the physics behind this analytic continuation}. For $d>2$,  the Riemann zeta function is connected with the sum~\eqref{eq:defSum} of evaluated conformal primary wavefunctions  on the chain~\eqref{eq:chain} of the boundary. For $d=1$, this sum is divergent and meaningless. The Riemann zeta function imposes us the power to bypass the divergence of this sum, but is there a deeper physics picture or physics concept for this analytic continuation on spacetime dimension of this sum? 

Another question is \emph{why the celestial circle $d=1$ is so special}, bearing the nontrivial zeros of the Riemann zeta function. On the celestial circle, the chain~\eqref{eq:chain} is a special discretization of the space, so it can not  shrink to zero and has nontrivial homology group. On higher integer dimensions $d\geq 2$, for example the celestial sphere $d=2$, the chain~\eqref{eq:chain} as a special discretized cycle can shrink to zero and so has trivial homology group. This suggests that the specialty of the celestial circle $d=1$ is connected with topology, when considering its connection with nontrivial zeros of the Riemann zeta function. 

Our last question is \emph{what is the meaning of the chain}~\eqref{eq:chain}. In the above paragraph it is explained as a discretized cycle in the boundary manifold, but that is only a special choice of the chain from fixing a bulk point $X$. There can be many choices of the chain, as long as we stick to the condition $-q(\vec{w}) \cdot X = n$ when defining the sum~\eqref{eq:defSum}. If we fix a boundary point $q(\vec{w})\in \mathbb{R}^{d}$, the resulting chain is a discretized cycle in the bulk. Instead of fixing a point, we can also choose an infinite number of pairs of points $(q_n, X_n)$ simultaneously from the bulk and the boundary, then the chain will be a special map of points connecting the bulk and the boundary. This suggests that when considering the specialty of the celestial circle $d=1$, we also need to take into account the connection with the bulk, that is, the holography itself. 

This consideration of the choices of the chain leads to a further consideration of the specialty of $d=1$. The bulk physics is a $D=2$ space with Minkowski time. If we select the pairs of points $(q_n, X_n)$ in such a way that on the bulk $\{X_n\}$ constitutes a square lattice, there might be a way to connect it with the $2D$ Hubbard model of strongly correlated condensed matter system. So this CCFT in $d=1$ might turn out to be a holographic description of the Hubbard model. We state here that this is only a possibility, and its chance could be nearly zero. 

In a word, our analysis connects the nontrivial zeros  of the Riemann zeta function  with the CCFT on the celestial circle. The number theory and topology of pure mathematics might play an important role in this connection. We hope that these results can help the development of new ideas in   related fields.

\begin{acknowledgments}
    Wei Fan is supported in part  by the National Natural Science Foundation of China  (Grant No.12105121).  
\end{acknowledgments}

\bibliography{riemann-zeta-ccft}%
\bibliographystyle{apsrev4-2}

\end{document}